# Effect of the applied magnetic field on formation of complex polyaniline films

Oleg P Dimitriev*, Petro M Lytvyn, Alla P Dimitriyeva and Ostap M Getsko

Address: Institute of Semiconductor Physics, National Academy of Sciences of Ukraine, Pr. Nauki 45, Kiev 03028, Ukraine

Email: Oleg P Dimitriev* - dimitr@isp.kiev.ua; Petro M Lytvyn - plyt@isp.kiev.ua; Alla P Dimitriyeva - alla_dimitriyeva@mail.ru; Ostap M Getsko - vlozovsk@isp.kiev.ua

* Corresponding author

## Abstract

Formation of complex polyaniline (PANI) films with diamagnetic TCNQ and paramagnetic metal ion impurities cast under applied magnetic field was studied. It has been found that the applied magnetic field affects interaction of PANI chains with the impurities and induces formation of magnetically ordered regions in the complex film doped by paramagnetic metal ions in contrast to the reference film of the same composition but prepared under ambient conditions. The magnetically ordered regions have been observed directly by scanning magnetic force microscopy. It was found a correlation in distribution of the magnetically ordered regions and peculiarities of the surface relief of a film. Electronic absorption spectra and conductivity measurements showed that an applied stationary magnetic field can suppress the interaction of PANI chains and paramagnetic metal ions and lowers conductivity of the resulting complex film up to one order of magnitude as compared with the reference film. An alternating magnetic field was found to improve interaction of PANI and diamagnetic TCNQ molecules. The mechanisms of the magnetic field influence on the complex film formation are discussed.

## Introduction

Effect of an applied magnetic field on chemical reactions, growth of organic films and functional state of organic materials is a subject of continuing interest of scientists in different areas of science at both fundamental and practical levels [1-7]. Despite many controversial results in this field because of usually small effects the magnetic field renders on growth and properties of functional films, some progress in understanding of how an applied magnetic field can control physical and chemical properties of organic materials has been achieved. It is known that the magnetic field can influence the orientation of a molecule through rotation of molecular frag-ments possessing a magnetic moment to adjust it to the applied field direction or renders the effect of translational motion of a molecule bearing a magnetic moment along the gradient of the applied field [8]. Since many organic functional materials have benzene and/or heterocyclic rings the applied magnetic field will induce an electric current inside the ring with the induced magnetic moment antiparallel to the applied field vector; thus the applied field will interact with the aromatic rings which act as diamagnetic dipoles. The above effects, however, can be noticeable when the magnetic forces overcome those of intermolecular interactions. So, it is expected that the effects can be observed either under a strong magnetic field or when the film is formed from a liquid or gaseous phase. Particularly, it was found that the magnetic field can induce an orientational effect on some organic polymers and biological molecules [9], as well as it can enhance mass transfer and morphological changes in organic films via magnetomechanical effect [10,11].

The effect of strong and stationary applied magnetic fields on formation and properties of polyaniline (PANI) has also been studied in the literature [12-22]. In the majority of these works two aspects of the applied magnetic field are considered, i.e., how strong magnetic fields of the order of 0.4–10 T affect chemical polymerization or electropolymerization of aniline, and how strong magnetic fields affect properties of PANI molecules with liquid crystalline groups or rare-earth dopants. Particularly, it was found that the applied magnetic field effects on aniline polymerization [15,16,19] and on specific morphology of the resulting PANI particles [12,14], that PANI molecules can be oriented in a solution [22], and that PANI films polymerized under a strong magnetic field applied along the film surface become anisotropic and show improved electroactivity and conductivity [18,20,21].

We have recently developed novel conducting complexes based on emeraldine base (EB) of PANI and tetracyanoquinodimethane (TCNQ) [23,24] as well as PANI and transition metal salts [25-27]. Particularly, it was shown that coordination of metal ions to the polymer chain serves two purposes. The first one is doping of the polymer chains, i.e., conversion of the insulating polymer form to the conductive form. The second purpose is directed to induce a specific morphology of the film. Since many transition metal ions possess paramagnetic properties, the applied magnetic field can better control formation of PANI-transition metal complexes, acting on both magnetic moments of metal ions and diamagnetic benzenoid or quinoid groups of PANI molecules. It is suggested that more weak applied magnetic fields are sufficient to affect the resulting film properties formed from the complex molecules bearing spins. This work is a systematic study of the effect of modest magnetic fields, not more than 0.1 T, both stationary and alternating, on morphology and properties of complex PANI films cast from solutions. This study aims to reach a better understanding of mechanisms of the applied magnetic field on organization of complex polymer molecules upon their condensation from a solution to a film.

## Experimental

### Sample preparation

Powder of emeraldine base (EB) of polyaniline (PANI) was dissolved in N-methylpyrrolidinone (NMP) to prepare a stock solution with concentration of 1 wt.%, followed by treatment of the solution in the ultrasonic bath and filtering of the solution to remove undissolved particles. Salts of $CoCl_2$, $Eu(NO_3)_3$, as well as tetracyanoquinodimethane (TCNQ), were dissolved in NMP to prepare solutions with desirable concentrations. Mixtures of EB and $CoCl_2$, EB and $Eu(NO_3)_3$, and EB and TCNQ were prepared by addition of the salt or TCNQ solutions to the EB solution with certain proportion of EB (calculated per tetramer units) to metal ions or TCNQ. The mixture usually consisted of 40 ml of the EB stock solution and 10 ml of the salt or TCNQ solution of concentration of $10^{-1}$ M to provide an approximate ratio 1:2 of a tetramer unit of the EB to metal cations or TCNQ.

The solutions were used for preparation of films by drop-casting onto glass plates. One series of samples was prepared under ambient conditions, and the other series of samples was prepared under the same temperature and atmospheric conditions but using an applied magnetic field. In the case when a stationary magnetic field was used, the sample on a thin glass substrate was placed in between two stationary magnets, with the opposite poles separated by about 0.5 cm and creating a magnetic field perpendicular to the sample surface with the strength of about ~60 mT in the area of the sample. The surface area of the magnetic poles exceeded that of the sample and the linear size of the poles was much larger the separation distance between the poles, so that the field between the poles was considered to be homogeneous.

In the case of applying an alternating magnetic field, a substrate with the sample was placed onto the solid surface under which a piece of a stationary magnet has been installed. The magnet had a magnetic axis oriented parallel to the substrate surface and a strength measured at the substrate surface of about ~15 mT. During evaporation of the solvent from the drop-cast film, the magnet was rotating around the axis which was perpendicular to the substrate surface with a frequency of several Hertz, so that the magnetic field vector laid mainly at the substrate surface but changed its direction continuously.

### Measurements

Magnetic force microscopy (MFM) measurements were performed by Dimension 3000 NanoScope IIIa scanning probe microscope. Magnetic force gradients were measured using frequency modulation at LiftMode operation (i.e., topography was scanned at the first pass in the tapping mode and magnetic field gradient was scanned at the second one using oscillation frequency shift of the magnetic probe moving at a constant height over surface). The value of lift scan height was about 100 nm; this value was chosen to provide fine resolution of MFM image and not to influence sharp surface features. The Veeco magnetic force etched silicon probes (MESP) with coerciv-ity of ~400 Oe and magnetic moment of $1 \cdot 10^{-13}$ Electro-Magnetic Unit (EMU) were used. Nominal tip radius was 25 nm.

Optical microscopy studies of the film morphology have been performed using a microscope "Axiostar plus" (Carl Zeiss) equipped with a photo-camera and a computer. Electronic absorption spectra were recorded using a dual-beam spectrophotometer "Specord M40" in the transmission mode, bare glass plate served as a reference. IR absorption spectra of the samples prepared on the germanium wafers were recorded using a dual-beam spectrophotometer "Specord M80".

The electrical in-plane conductivity of the films was measured using a standard four-probe technique. During the measurement, an appropriate constant current, $I$, in the range 0.1 – 10 µA was maintained on two outer probes, and the voltage drop, $V$, was measured across two inner probes, using a UNI-T M890C$^+$ electrometer. The resulting conductivity, $\sigma$, was found according to the expression for the thin-film approximation [28], $\sigma = \ln2 \, (I/\pi \, d \, V)$, where $d$ is the film thickness.

## Results and discussion
### Effect of magnetic field on formation of pristine PANI films

It has been found that both stationary and alternate magnetic fields applied to EB films cast from NMP solutions at room temperature do not substantially affect the film morphology. Both films prepared under applied magnetic fields and under ambient conditions at room temperatures showed approximately the same morphology. A negligible effect of the alternating magnetic field observed here in respect to EB films is in contrast to our previous results which showed that other organic films prepared under alternating magnetic field at room temperatures change their morphology significantly [11]. The negligible effect for the EB films can be due to relatively high viscosity of the polymer solution and small mobility of the polymer chain as compared to low-weight organic molecules studied in ref. [11]. However, a small increase of the processing temperature to about ~50°C resulted in noticeable morphological changes of the EB films, i.e., EB films prepared under alternating magnetic field showed more smooth morphology consisting of enlarged homogeneous regions with small dense islands in contrast to the reference film whose morphology was highly inhomogeneous (Fig. 1). Such a result is consistent with those observed for other organic films formed under an alternating magnetic field [11].

An IR absorption spectrum of the EB film cast under applied alternating magnetic field was different from the spectra of films prepared at ambient conditions or under a stationary magnetic field. We could not observe an effect of the applied stationary magnetic field on the IR absorption spectrum. On the other hand, the alternating magnetic field acting at the increased temperature, induced changes in the IR spectrum. IR spectra of EB films prepared under alternating magnetic field and without it at ~50°C are compared in Fig. 2. The first difference is that the absorption bands of the NMP solvent at 1680 and 1648 cm$^{-1}$ are more pronounced in the film prepared under alternating magnetic field. That means that the solvent was entrapped into the film and therefore plasticizes it to provide a smother morphology as compared to the reference film. The second significant change was observed for the absorption at ~1300 cm$^{-1}$ which consists of the bands at ~1328 cm$^{-1}$ and ~1285 cm$^{-1}$. As known, the absorption band at ~1285 cm$^{-1}$ in EB is normally assigned to the C-N stretching vibration [29], whereas that at 1328 cm$^{-1}$ to the radical-cation C-N$^{\cdot+}$ stretching vibration observed normally in the doped PANI chains [30-32]. A more pronounced band at 1328 cm$^{-1}$ observed for the film prepared under alternating magnetic field testifies to a higher protonation level of the polymer chains in this case. This can be assigned to higher overlap and interactions of polymer chains in this film, when the interaction between the chains leads to the radical-cation formation (protonated imine groups) as a result of proton exchange between the polymer chains. Such a hypothesis is supported by observation of paramagnetic spins with the Pauli-like behavior in even carefully synthesized emeraldine base of PANI [33], which was explained by that the polarons in EB are created in pairs and that the intrapair interactions encompass a broad distribution of antiferromagnetic exchange couplings [33,34].

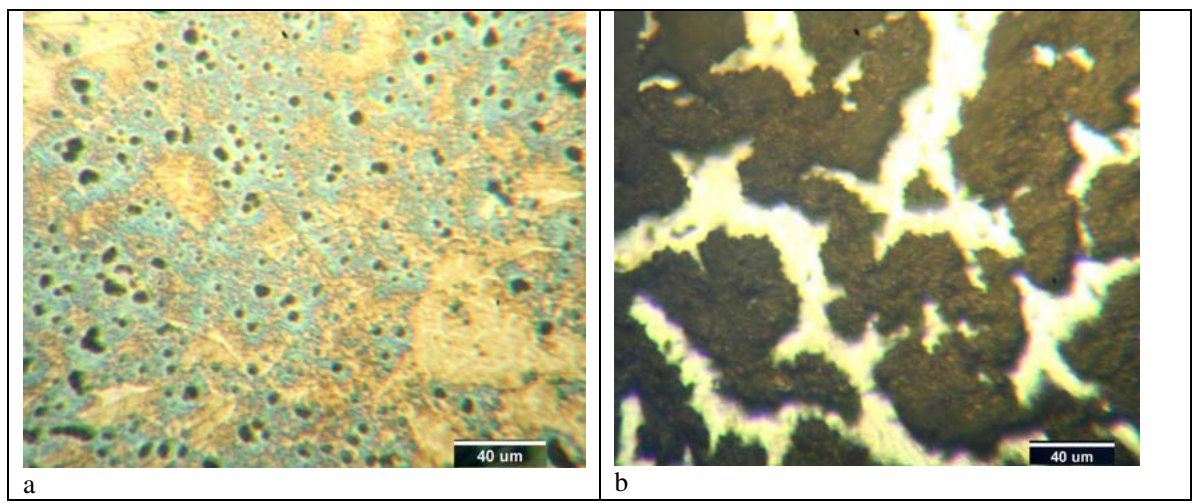

**Figure 1**
Optical microscopy images of EB films cast (a) under alternating magnetic field at ~50°C and (b) without applied magnetic fields at ~50°C.

An electronic absorption spectrum of the films yields an additional evidence of the effect of the radical-cation formation (Fig. 3). In the film prepared under alternating magnetic field at ~50°C, an absorption band at ~630 nm which corresponds to the benzenoid-to-quinoid electronic transition [35] was lowered and red-shifted, while the absorption in the near-IR increased. The latter is indicative of polarons whose origin can be, probably, due to intrachain proton or electron transfer resulting in formation of NH$^{+\cdot}$ radical-cations in the polymer backbone [36].

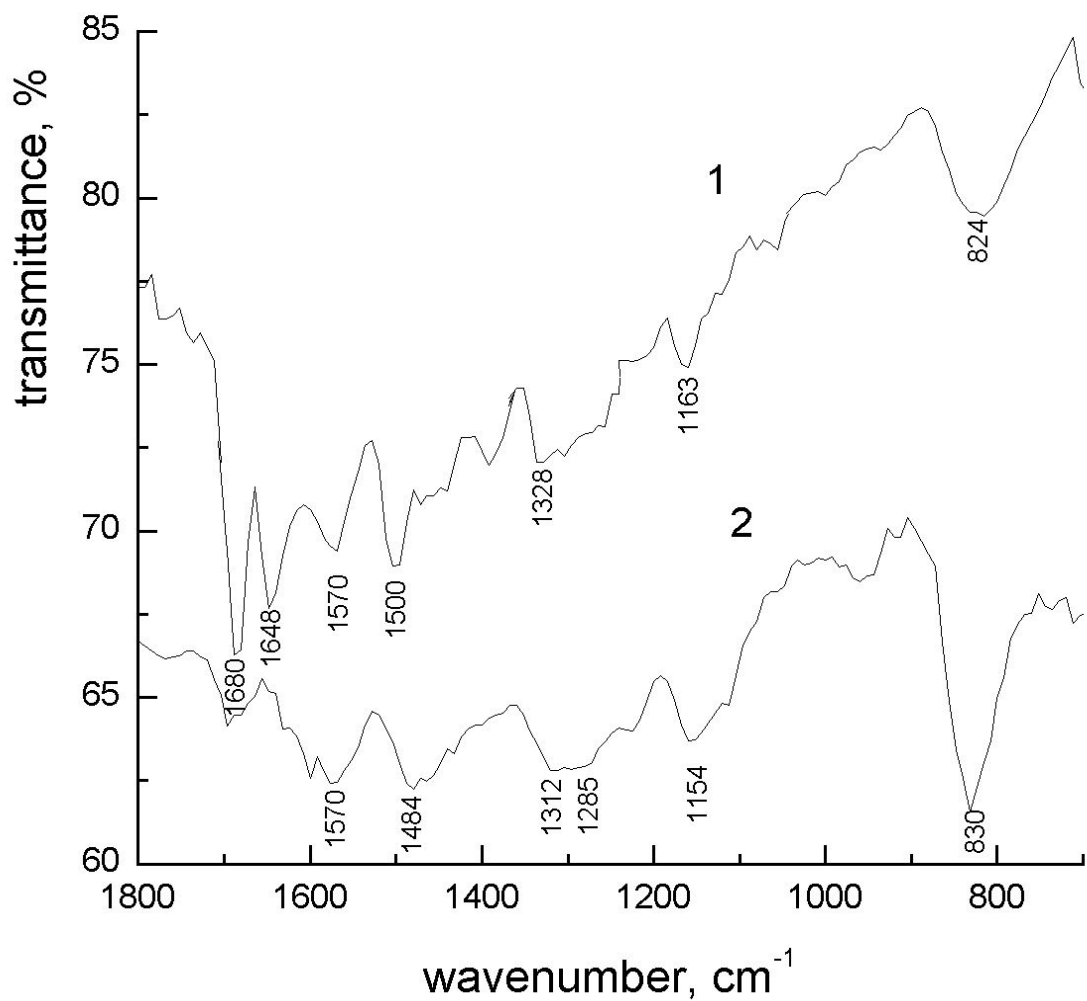

**Figure 2**
IR absorption spectra of the EB films cast (1) under alternating magnetic field at ~50°C and (2) without applied magnetic fields at ~50°C.

Thus, one can suggest that an applied magnetic field facilitates a proton exchange between chain fragments in highly entangled chains or between the polymer backbone and the residual solvent molecules in thin film with corresponding formation of polarons on the polymer chain. This process should be controlled by the N-H...N intermolecular hydrogen bonding whose energy is of the order of 0.1 eV [37]. A crude estimation shows that to reach such an energy level the magnetic field should be several orders of magnitude higher than that used in our experiments because it creates the energy density of only ~$10^{-6}$ eV per a hydrogen bond (assuming that the polymer has one hydrogen bond per a tetramer unit and that the polymer chains are closely packed). However, the weak external influence of the magnetic field can be quite effective when there is an additional factor affecting the hydrogen bonding, for example, thermal motion of the chains. That is why the above effects were observed only at elevated temperature and upon a slow formation of films from a solution. Moreover, the intramolecular proton migration which leads to interchange of the benzenoid and quiniod groups in the polymer backbone requires smaller energy contributions and therefore it is more affected by the external magnetic field.

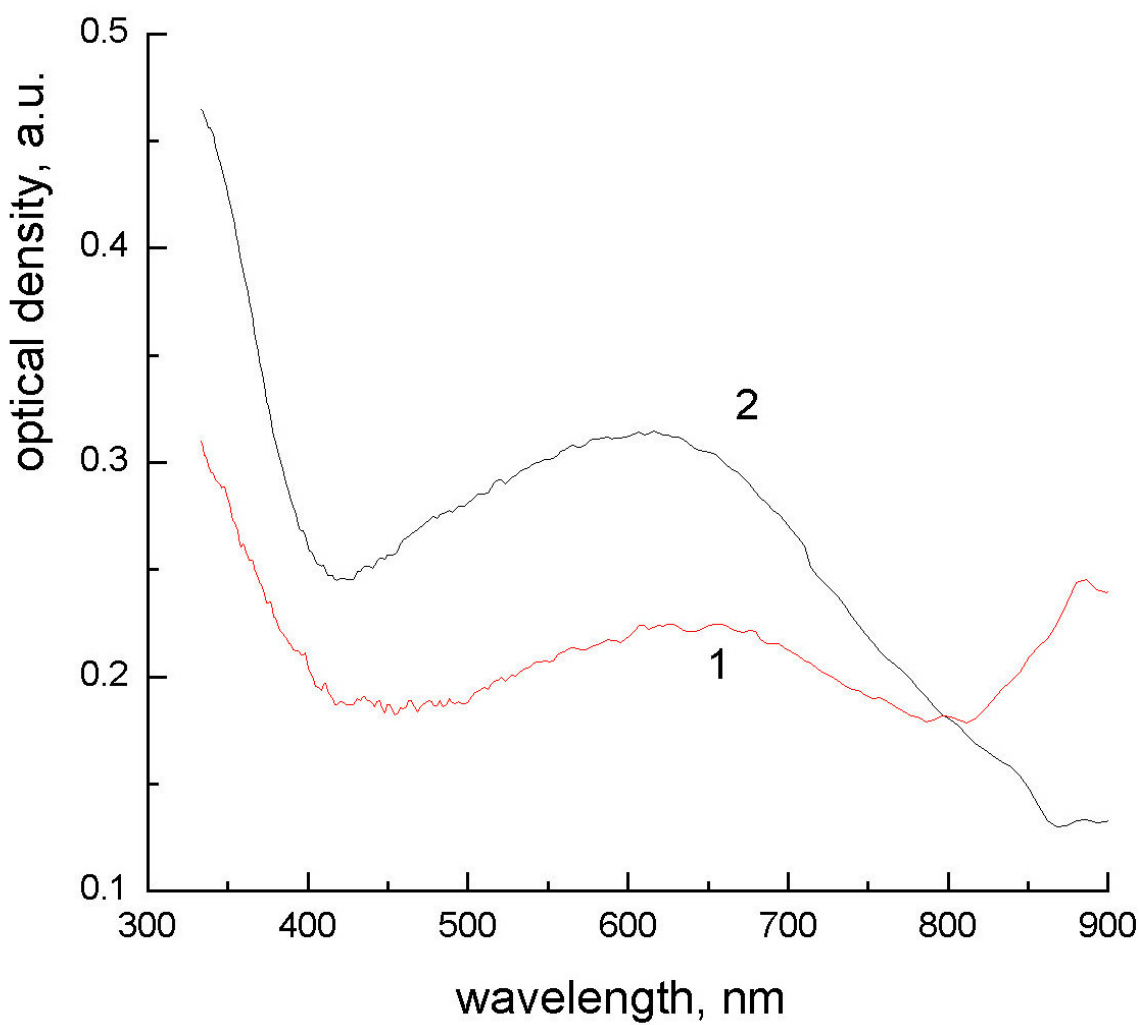

**Figure 3**
Electronic absorption spectra of EB films cast (1) under alternating magnetic field at ~50°C and (b) without applied magnetic fields at ~50°C.

### *Effect of magnetic field on formation of surface relief of the complex films*

As was shown in our previous works [26,27], the morphology of complex PANI films is greatly dependent on the transition metal salt used. Particularly, it has been shown that EB films doped by chlorides of metals of the ferrous group (Fe, Ni, Mn) have, as a rule, a grained morphology. In a similar way, it was observed in this study that salt of $CoCl_2$ induces a similar grained morphology of the complex $EB:CoCl_2$ film (Fig. 4a). The morphology of $EB:CoCl_2$ films prepared both under ambient conditions and under the stationary magnetic field was found to be similar. The morphology of films prepared under alternating magnetic field even at room temperatures was noticeably different from the above cases. AFM images showed larger structural units in the form of hillocks in the film formed under alternating applied field as compared to the film prepared under ambient conditions or stationary magnetic field (Fig. 4). Such a difference is consistent with the concept of magnetomechanical forces acting on molecules in the course of the film formation under alternating magnetic field and resulting in many cases in circular structures [11]. An alternating magnetic field, of course, is always accompanied by an alternating electric field through the induction effect, which can influence behavior of the polymer segments possessing an electric dipole. We cannot distinguish the contribution of the magnetic and electric component of the electromagnetic field in formation of the film morphology in this case. However, the changes in the magnetic structure of the film probed by the MFM technique can help us to understand whether the magnetic field really influences the film formation.

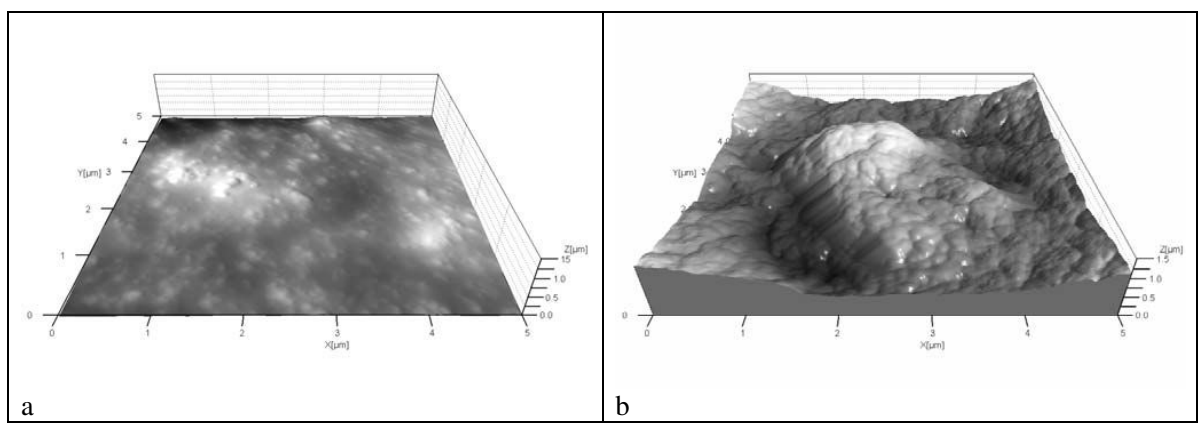

**Figure 4**
AFM images of EB:CoCl$_2$ films prepared (a) under stationary magnetic field and (b) under alternating magnetic field.

### *Effect of magnetic field on magnetic ordering in the complex films*

Using the magnetic force microscopy, it has been found that the films of EB:CoCl$_2$ prepared under ambient conditions do not have magnetically ordered regions within the resolution scale of the MFM method. On the other hand, films prepared under applied magnetic field, both stationary (normal to the sample surface) and alternating, showed regions with the magnetic ordering. These regions are indicated in Fig. 5b by the dark and white color and probably correspond to magnetic domains with anti-parallel magnetization of these domains, respectively. It can be seen that the magnetic domains are randomly scattered over the film surface. However, the comparison of the AFM and MFM images reveals correlation in distribution of the topographical and magnetic features (Figs. 5a and 5b). The most surprising fact is that the magnetic domains of the opposite signs are formed within the hillocks and cavities of the surface relief, respectively. Fig. 5 shows the correlation of the topographical relief and magnetized domains for a randomly chosen region on the film surface. Again, such a correlation can be explained by magnetomechanical forces acting on the material and enhanced in those regions where magnetization is higher. So, the higher is the magnetization, the stronger is the change in the surface morphology due to magnetomechanical forces.

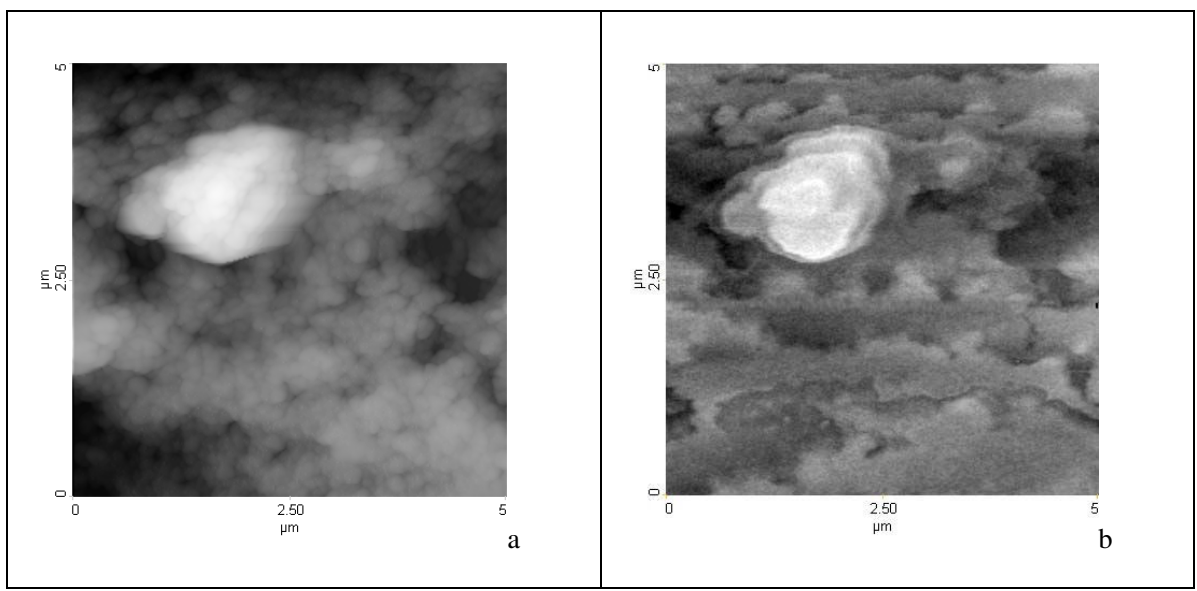

**Figure 5**
Comparison of (a) AFM and (b) MFM images of the same region of the EB:CoCl$_2$ film prepared under alternating magnetic field. Gray scale colors correspond to 1.6 μm and 2.9 Hz for AFM and MFM images, respectively.

### *Interaction of PANI molecules and paramagnetic metal ions in the presence of applied magnetic field*

In addition to the influence on the film morphology, an applied magnetic field was found to affect other physical properties such as interaction of the polymer and the metal ions which play a role of dopants to increase film conductivity. This effect was found to depend on kinetics of solvent evaporation during film preparation. Relatively fast kinetics of evaporation did not result in any significant difference between films formed under stationary magnetic field and films formed under ambient conditions, respectively. The electronic absorption spectra of the films were similar and conductivity varied within 20% from one sample to another within each series of experiments. Slow formation of complex PANI films from evaporated polymer solutions normally resulted in some phase separation of the unaffected EB and EB doped by metal ions, respectively. It has been found that a slow formation of complex PANI films from evaporated polymer solutions under the stationary magnetic field of about 60 mT applied normally to the film surface leads to the changing size and redistribution of polymer regions doped by metal ions. Fig. 6 shows such a distribution of green and blue regions in the EB:Eu(NO$_3$)$_3$ film, where green color corresponds to the doped conducting polymer phase, while blue color corresponds to the unaffected insulating EB phase. It can be seen that the film formed under applied magnetic field consists of larger clusters as well as it has larger area of blue regions, i.e., it is less conductive. Direct measurements of conductivity of the films by the four-probe technique confirmed that the conductivity of the EB:Eu(NO$_3$)$_3$ film prepared under ambient conditions is about one order of mag-nitude higher as compared to its counterpart film prepared under the applied magnetic field, being 1.2 × 10$^{-4}$ and 2.0 × 10$^{-5}$ Scm$^{-1}$, respectively.

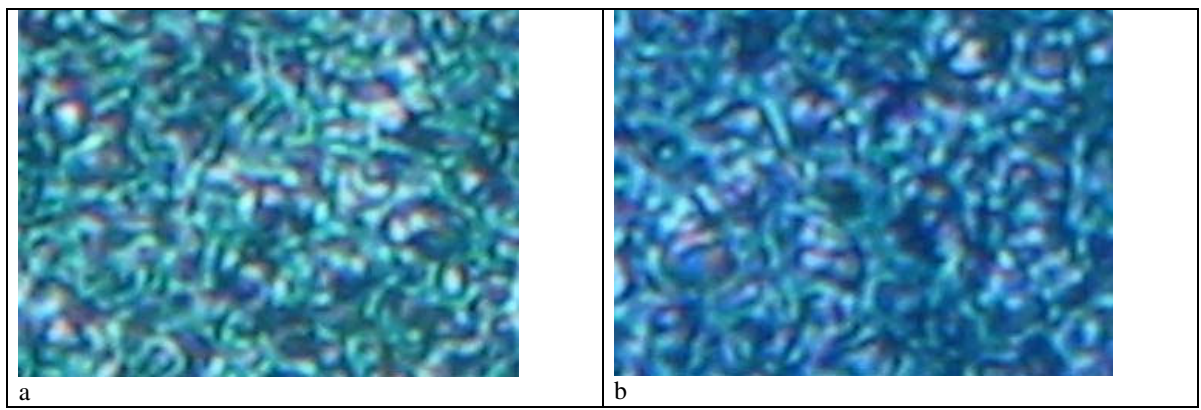

**Figure 6**
Optical microscopy images of EB:Eu(NO$_3$)$_3$ films prepared (a) under ambient conditions and (b) under applied magnetic field normal to the surface. Dimensions of the images are 33 × 24 μm$^2$.

Images of the EB:CoCl$_2$ films showed the analogous features, i.e., the complex film prepared under applied magnetic field had larger structural units and reduced features of doping by Co$^{2+}$ ions as compared to its counterpart film prepared under ambient conditions (Fig. 7). A higher impact of Co$^{2+}$ ions is seen in Fig. 7a through the change of the film color from blue to violet and thicker network of green veins corresponding to the conductive pathways.

The changed ability of doping of PANI chains by the Co$^{2+}$ and Eu$^{3+}$ ions under applied magnetic field can be due to available magnetic moment of these ions. Doping of PANI usually results in appearance of radical nitrogen ions on the polymer chain which are paramagnetic. Interaction of paramagnetic metal ions and paramagnetic nitrogen ions is unfavorable in the presence of the applied magnetic field, because anti-parallel orientation of interacting spins will be suppressed. Thus, the applied magnetic field is unfavorable for doping of the EB chain by paramagnetic metal ions.

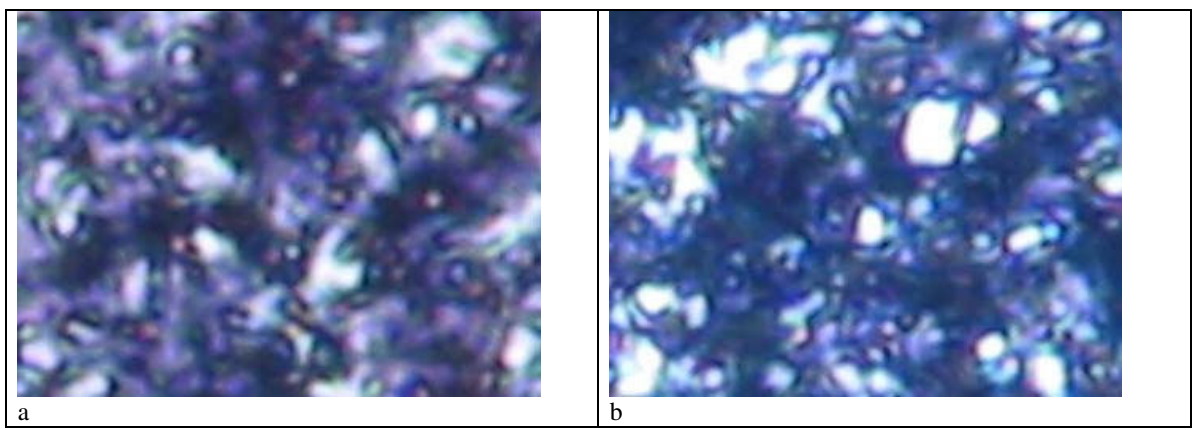

**Figure 7**
Optical microscopy images of EB:CoCl$_2$ films prepared (a) under ambient conditions and (b) under applied magnetic field normal to the surface. Dimensions of the images are 33 × 24 μm$^2$.

### *Effect of magnetic field on formation of PANI:TCNQ complex films*

TCNQ is a diamagnetic molecule due to the negative magnetic susceptibility of its quinoid moieties [38]. As has been shown previously, TCNQ can interact with the EB polymer via charge-transfer interaction, accepting an electron from the electron-rich nitrogen atom of the amine group of EB and leading to formation of anion-radical of $TCNQ^{\bullet-}$ and cation-radical of EB, respectively [23]. The resulting complex film has also the changed electrical and optical properties [24]. Interaction of the EB and TCNQ can also change the diamagnetic status of the EB:TCNQ complex if spin unpairing of the resulting radicals takes place in the complex molecule.

Electronic absorption spectra can distinguish the neutral and anion-radical forms of TCNQ. There are normally three absorption bands typical of the monomeric radical form of TCNQ, exhibited at ~420 nm, ~760 nm and ~850 nm [39,40]. These absorption bands can be found in spectra of cast TCNQ films (Fig. 8), the first one as a shoulder on the low energy side of the major band, and the other two are merged into a single IR band centered at 843 nm. There is one more band in spectra at ~658 nm which can be attributed to biradicals $(TCNQ^{\bullet-})_2$ [39]. The presence of anion-radicals of TCNQ in the condensed film can be explained by the fact that TCNQ can be ionized when dissolved in a polar solvent [23] and that the TCNQ anions can be entrapped into the cast film. Fig. 8 provides the evidence that the alternating magnetic field applied to the cast film leads to higher amount of anion-radicals present in the film as compared to the film prepared under ambient conditions. A similar evidence can be derived from the electronic spectra of the complex EB:TCNQ films (Fig. 9), where the absorption of anion-radicals of TCNQ still prevails in the complex film prepared under alternating magnetic field as compared to the complex film prepared under ambient conditions. These results also support the idea that the applied magnetic field favors formation of radicals rather than neutral molecules in the complex films.

Comparison of electronic absorption spectra of EB:TCNQ films prepared under different conditions shows that the absorption band of EB at 625 nm is seen only in the film prepared at ambient conditions. This result indicates that some phase separation of EB and TCNQ takes place in the film. Microscopy images (Fig. 10) confirm this conclusion. Blue region in Fig. 10a corresponds to the EB phase of the film, whereas the yellow region to the TCNQ phase. On the other hand, the film prepared under alternating magnetic field showed a homogeneous color over the whole film area (Fig. 10b). Therefore, one can conclude that the applied magnetic field promotes interaction of the EB and TCNQ molecules via charge-transfer. It should be noted that the microstructure of the resulting films is also different and depends on the preparation conditions used.

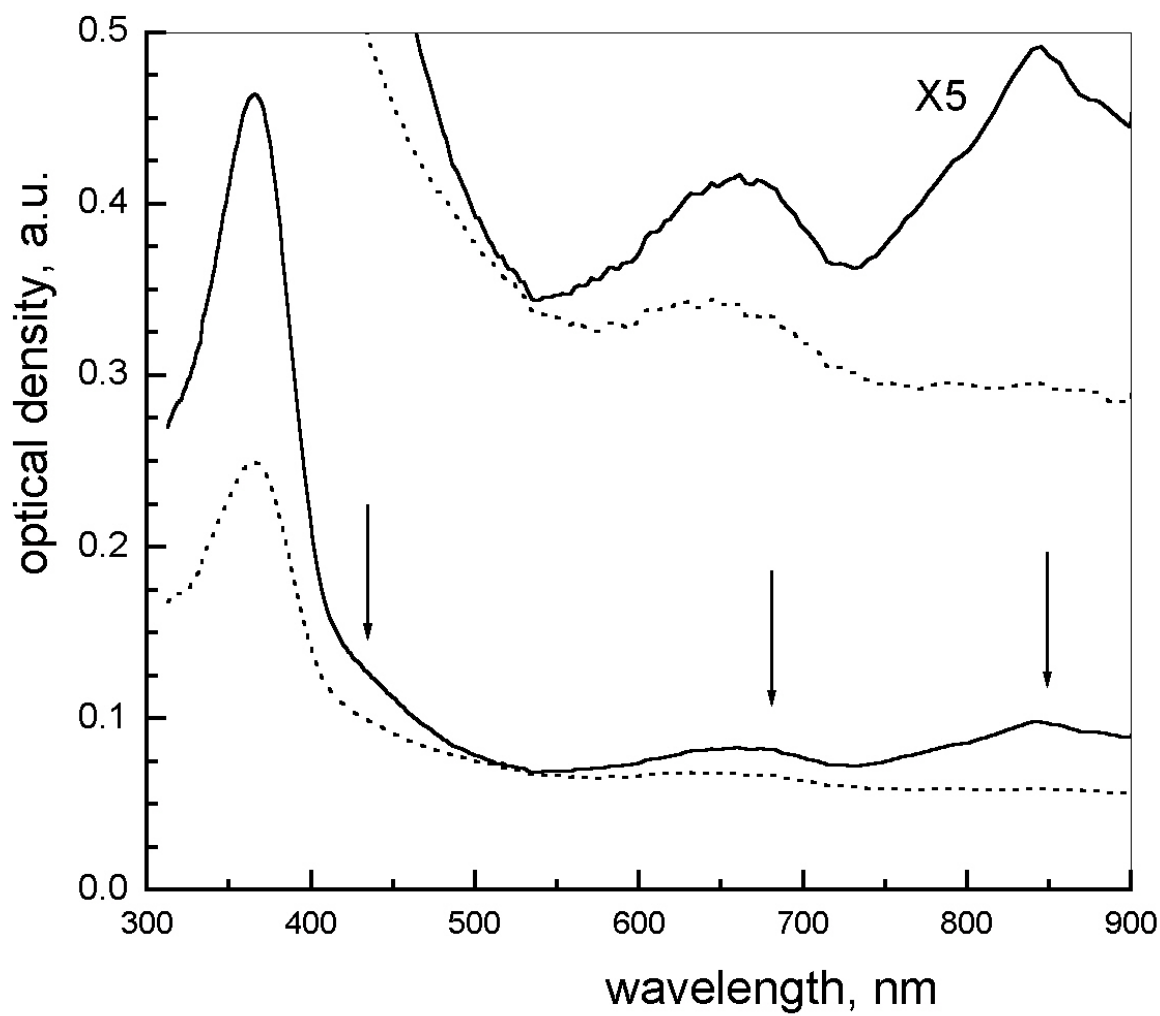

**Figure 8**
Electronic absorption spectra of TCNQ films cast at ambient conditions (dotted curve) and under alternating magnetic field (solid curve). Arrows indicate the absorption bands attributable to the radical-ion formation of TCNQ.

The film cast under alternating magnetic field shows larger structural units as compared to the film obtained at ambient conditions (Fig. 10).

## Conclusion

A moderate magnetic field was found to influence formation of both pristine and complex PANI films, leading to the changed film morphology and physical properties. It has been found that the alternating magnetic field affects greatly the morphology of the complex films which was explained by magnetomechanical effect, i.e., a motion of magnetized particles or diamagnetic molecular fragments in the presence of the applied magnetic field. The electric component of the alternating electromagnetic field can probably contribute to this effect also. However, the observed correlation in distribution of topographical features and magnetically ordered regions on the film surface testifies to significant contribution of the magnetic field component to this effect. Therefore, one can conclude that the influence of the applied magnetic field on the film morphology can be enhanced via the presence of paramagnetic impurities in the film structure.

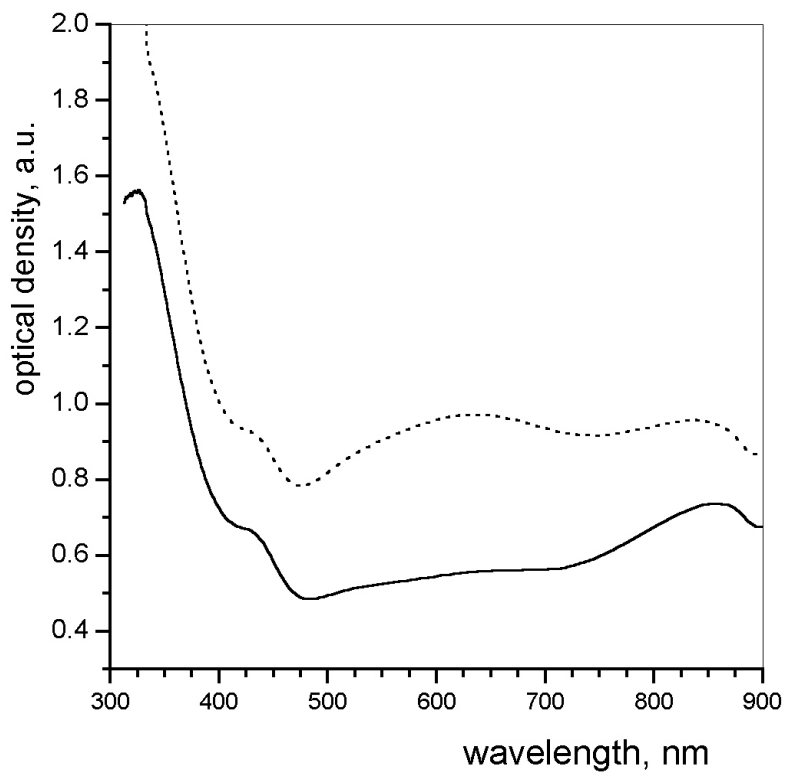

**Figure 9**
Electronic absorption spectra of EB:TCNQ films cast at ambient conditions (dotted curve) and under alternating magnetic field (solid curve).

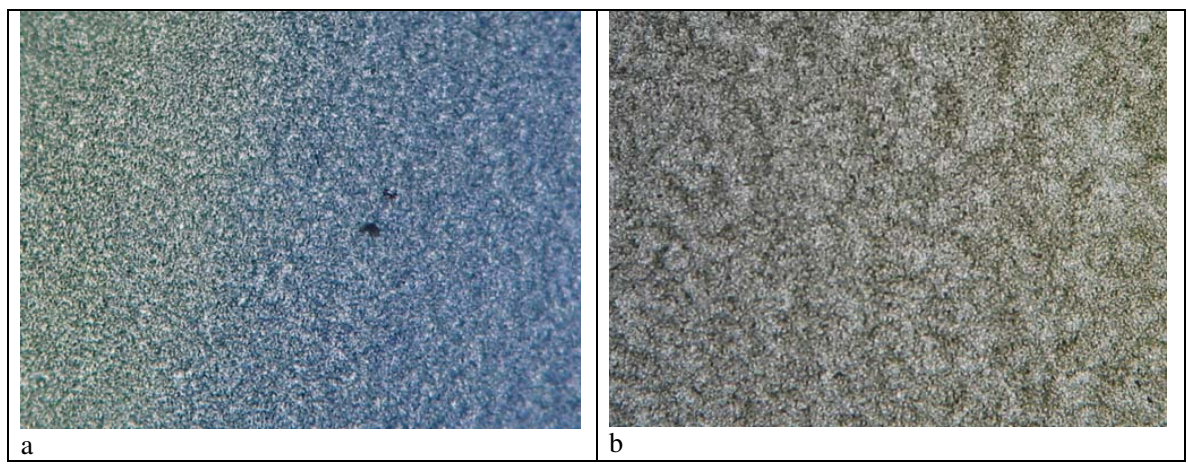

**Figure 10**
Optical microscopy images of EB:TCNQ films prepared (a) under ambient conditions and (b) under alternating magnetic field. Dimensions of the images are 660 × 480 $\mu m^2$.

It has been found that the applied magnetic field tends to increase the amount of radicals on the polymer chain or in the PANI-TCNQ charge-transfer complex, i.e., to induce proton or electron exchange, respectively. We speculate that the origin of this effect consists in that the applied magnetic field supports the existence of radical molecular fragments formed spontaneously upon thermal motion and intermolecular interaction during film formation. Two different effects of the complex film formation owing to the suggested mechanism have been observed. The first one, when one of the components bears unpaired spin, is the suppressed interaction of EB and paramagnetic metal ions leading to the phase separation and decreased conductivity of the resulting complex films. The second effect, when both counterparts have paired spins, is the improved interaction of the EB and TCNQ leading to the electron exchange and formation of homogeneous complex films. However, further experiments to probe spin amount in the samples prepared under applied magnetic field are necessary to clarify the observed effects in more detail.